\documentclass[apj]{emulateapj}
\usepackage{graphicx}
\usepackage{color}
\usepackage[titletoc,toc,title]{appendix}
\usepackage{threeparttable}

\shorttitle{High resolution BH fueling in a merger system}
\shortauthors{Prieto et al.}

\begin{document}

\title{Black hole fueling in galaxy mergers: a high-resolution analysis}

\author{Joaquin Prieto$^1$, Andr\'{e}s Escala$^1$, George C. Privon$^2$ \& Juan d'Etigny$^1$}
\affil{$^1$Departamento de Astronom\'{i}a, Universidad de Chile, Casilla 36-D, Santiago, Chile.\\
$^2$Department of Astronomy, University of Florida, 211 Bryant Space Sciences Center, Gainesville, FL 32611, USA.}

\begin{abstract}

Using parsec scale resolution hydrodynamical adaptive mesh refinement simulations 
we have studied the mass transport process throughout a galactic merger. 
The aim of such study is to connect both the peaks of mass accretion rate onto 
the BHs and star formation bursts with both gravitational and 
hydrodynamic torques acting on the galactic gaseous 
component. Our merger initial conditions were chosen to mimic a realistic system.
 The simulations include gas cooling, star formation, supernovae feedback, 
and AGN feedback. Gravitational and hydrodynamic torques near pericenter passes 
trigger gas funneling to the nuclei which is associated with bursts of star 
formation and black hole growth. Such 
episodes are intimately related with both kinds of torques acting on the 
galactic gas. Pericenters 
trigger both star formation and mass accretion rates of $\sim$ few $(1-10)\,M_\odot$/yr. 
Such episodes last $\sim\,(50-75)$ Myrs. Close passes also can 
produce black hole accretion that approaches and reaches the Eddington rate,
lasting $\sim$ few Myrs. Our simulation shows that both gravitational 
and hydrodynamic torques are enhanced at pericenter passes with gravitational torques tending to have higher values than the hydrodynamic torques throughout the merger. We also find that in the 
closest encounters, hydrodynamic and gravitational torques can be comparable in their effect on the gas, the two helping in the redistribution of both angular momentum 
and mass in the galactic disc. Such phenomena allow inward mass transport onto 
the BH influence radius, fueling the compact object and lighting up the galactic 
nuclei.

\end{abstract}


\section{Introduction} 

Cosmological N-body numerical simulations of structure formation shows that 
dark matter (DM) haloes were formed by mergers between smaller haloes in 
a hierarchical way \citep[e.g. ][]{Angulo+2012}. Similar kinds of simulations 
including baryonic physics have shown that galaxies were formed inside DM 
haloes in this hierarchical model \citep[e.g. ][]{Dubois+2014,Genel+2014}. 
In this sense, mergers and interaction between galaxies are a fundamental 
piece of the galaxy formation process and certainly they influence the 
galaxies evolution. In fact, observation of irregular and disrupted systems 
are consistent with mergers and interactions between galaxies 
\citep[e.g. ][]{TT72,Schweizer82,Engel+2010,Bussmann+2012}.  

The infrared and optical properties of interacting galaxies are different
compared with isolated systems \citep[e.g.][]{SandersMirabel1996}. Such 
differences can be a consequence of star formation burst associated to 
galactic interactions \citep[e.g.][]{Sanders+1988,Duc+1997,Jogee+2009}. 
Beside the radiative signatures of star formation, some 
interacting galaxies also show nuclear activity which can be associated with 
black hole (BH) fueling \citep[e.g.][]{Petric2011,Stierwalt2013}
These two features, i.e. SF bursts and active galactic nuclei (AGN), suggest 
that galactic encounters are able to redistribute gas inside galaxies, 
moving material toward their central regions to feed massive BHs and 
trigger SF bursts \citep[e.g.][]{BarnesHernquist1991,MihosHernquist1996,Springel+2005a}.

Smoothed Particle Hydrodynamic (SPH) numerical simulations of galactic 
mergers with $\sim$ few $(10-100)$ pc of resolution have shown that gravitational 
torques are able to produce inflows of gas toward the galactic central 
regions \citep[e.g.][]{BarnesHernquist1991,WursterThacker2013,NewtonKay2013,Blumenthal+Barnes2018}. 
Such inflows, increase the gas density of galactic centers enhancing SF 
\citep[e.g.][]{Teyssier+2010,Powell+2013} and at the same time are able to feed 
central super massive BHs triggering AGN activity 
\citep[e.g.][]{Sanders+1988,Bahcall+1995,Debuhr+2011}. Besides the low resolution 
SPH simulations including both SNe and AGN feedback mentioned above 
\citep[e.g.][]{Debuhr+2011,WursterThacker2013,NewtonKay2013}, using an adaptive 
mesh refinement (AMR) simulation of $\sim$ 8 pc of resolution \citet{Gabor+2016} 
have also shown that pericenter passes correlate with peaks in both BH and stellar 
activity but did not analyze the source of torques. The lack of parsec scale resolution AMR 
simulations studying mass transport in galaxy mergers strengthen the relevance 
of torque analysis in this kind of experiments.
However studies of mass transport in $\sim$pc-resolution simulations have not been performed. 

When choosing to deal with idealized mergers using AMR codes over a Lagrangian code, problems with the advection of material and grid alignment issues may arise, which could result in a loss of angular momentum conservation \citep{Wadsley+2008,Hahn+2010,Hopkins2015}. This issues are minimized as high spatial resolution is imposed at central galaxy regions, minimizing spurious field misalignments, and also, since pericenter passes have short durations (few orbital times at most) there are no significant orbital angular momentum deviations with respect to the ideal case. Furthermore, the galaxies are maintained at resolutions that are high enough for the AMR technique to be effective at resolving shocks throughout the merger process and therefore contact discontinuities are captured.

As the objective of our simulation is to properly and fully characterize a generic galaxy merger that exhibits realistic dynamics, we have to choose appropriate orbital initial conditions that  have  been  proven  capable  of  nearly-reproducing  such observed controlled environments.   Due  to the  degeneracy of the problem and the large parameter space of galaxy interactions, constraining the initial conditions with hydrodynamic simulations would be prohibitively time-consuming. Privon et al. (2013) used the Identikit code to find the orbital parameters capable to reproduce the morphology and kinematics of tidal features of four known observed galaxy mergers (NGC 5257/8, The Mice, Antennae and NGC 2623).  In this work we will adopt their orbital parameters (as an ansatz) for NGC 2623. Whilst the objective of this work will not be to reproduce neither the morphology of this system, we cite its characteristics as order of magnitude control values. 

NGC 2623 is a low-redshift, luminous infrared galaxy (LIRG) with an infrared luminosity of $L_{\rm IR}=3.6\times 10^{11}L_{\odot}$ from \citet{Armus2009}. The system 
has been classified as an M4 merger \citep{Larson+2016}, i.e. they are 
galaxies with apparent single nucleus and evident tidal tails. The merger shows two tidal tails of $\sim 20-25$ kpc 
in length, approximately with a single nucleus in IR \citep[e.g.][]{Evans+2008}. 
\citet{SandersMirabel1996} found a system stellar mass of 
$M_\star=2.95\times 10^{10}M_\odot$ with a molecular hydrogen mass of 
$M_{\rm H_2}=6.76\times 10^{9}M_\odot$. 
This values do not stray afar from typical LIRG values found in samples like the GOALS survey \citep{Armus2009}. \citet{Haan+2011} show that although there is some spread in the central BH mass values found in the GOALS survey, masses are generally found in the $10^7-10^9\,M_\odot$ range.

In this paper, for the first time we will study the evolution of a merger system from its early stages, up to the point where their BHs coalesce, using 
a $\sim$ 3 pc resolution AMR simulation including SF, supernovae (SNe) feedback, BH 
particles and AGN feedback. The goal of this paper is to understand the connection 
between torques, SF bursts and AGN activity in such large scale galactic 
environments with unprecedented high resolution, resolving the BH influence radius.

The paper is organized as follows. In \S 2 we describe 
the numerical details of the experiment, in \S 3 we show our results and in \S 4 
we present our discussion and conclusions.

\section{Methodology and Numerical Simulation Details}
\label{Methodology}

\subsection{Initial conditions}

As already mentioned, in this work we use the parameters found in Privon et al. (2013) for NGC 2623,
as initial conditions for a high 
resolution hydrodynamic numerical simulation. Table \ref{tab:t1} shows 
the initial orbital parameters of the simulated merger used in this work 
and table \ref{tab:t2} shows the initial position and velocity for both 
galactic centers. 

In addition to the orbital ICs, it is necessary to specify both 
the mass content and the mass distribution for each component of the galaxies,
including the gaseous disc, stellar disc, and stellar bulge. 
In order to create ICs for the DM haloes, gas and stars for each galaxy we 
have used the DICE code \citep{Perret+2014}. For our setup the gaseous disc follows an
exponential profile with a characteristic radius of 1 kpc. The stellar disc is modelled with a Myamoto-Nagai profile with a characteristic radius of 0.677 kpc. The stellar bulge
follows a Einasto profile with a characteristic radius of 0.6 kpc. Finally for the DM profile we employ a Navarro Frenk and White profile \citep{NFW1996} with a concentration parameter equal to 10. The SFR in the stellar disc follows \citet{Bouche+2010}. Table 
\ref{tab:t3} shows a complete summary of the galactic parameters of the system.

\begin{table}[h!]
\begin{center}
\caption{Initial orbital parameters.}
\begin{tabular}{c c c c c c c}
\hline\hline
$D_{\rm ini}$ [kpc]& $e$ & $p$ [kpc]& $\mu$ & $(i_1;\omega_1)$       & $(i_2;\omega_2)$\\
\hline 
50.0               & 1.0 & 0.6      &  1.0  & $(30^\circ;330^\circ)$ & $(25^\circ;110^\circ)$\\
\hline
\end{tabular}
\end{center}
\begin{tablenotes}
\item the orbit $e$, pericentral distance of the orbit $p$, mass galaxies ratio 
$\mu$ and disk orientation for both galaxies $(i,\omega)$ with respect 
the orbital plane.
\end{tablenotes}
\label{tab:t1}
\end{table}

\begin{table}[h!]
\begin{center}
\caption{Initial position and velocity.}
\begin{tabular}{l c c}
\hline\hline
 Coordinates                        & Gal$_1$             & Gal$_2$               \\
\hline 
(x, y, z) [kpc]                     & (-25,\,0,\,0)     & (25,\,0,\,0)        \\
(v$_x$, v$_y$, v$_z$) [km/s]        & (25,\,4.1,\,0.0)  & (-25,\,-4.1,\,0.0)  \\
\hline
\end{tabular}
\end{center}
\begin{tablenotes}
\item The orbital plane has been rotated 45$^\circ$ in the polar direction and
45$^\circ$ in the azimuthal direction. Those data are in the reference frame of
the simulated box.
\end{tablenotes}
\label{tab:t2}
\end{table}

\begin{table}[h!]
\begin{center}
\caption{Initial isolated galaxy setup.}
\begin{tabular}{l c}
\hline\hline
Gaseous disc                        &                   \\
(Exponential profile)               &                   \\
\hline
Mass [$10^{9}{\rm M}_\odot$]        & 1.0               \\
Characteristic radius [kpc]         & 1.0               \\
Truncation radius [kpc]             & 5.0               \\
\hline
Stellar disk                        &                   \\
(Myamoto-Nagai profile)             &                   \\
\hline
Mass [$10^{9}{\rm M}_\odot$]        & 2.975             \\
Number of particles                 & 11900000            \\
Characteristic radius [kpc]         & 0.677             \\
Truncation radius [kpc]             & 5.0               \\
\hline
Stellar bulge                       &                   \\
(Einasto profile)                   &                   \\
\hline
Mass [$10^{9}{\rm M}_\odot$]        & 0.975             \\
Number of particles                 & 390000             \\
Characteristic radius [kpc]         & 0.6               \\
Truncation radius [kpc]             & 1.0               \\
\hline
Dark mater halo                     &                   \\
(NFW profile)                       &                   \\
\hline
Mass [$10^{9}{\rm M}_\odot$]        & 20                \\
Number of particles                 & 200000             \\
Concentration parameter             & 10                \\
Truncation radius [kpc]             & 60                \\ 
\hline
\end{tabular}
\end{center}
\label{tab:t3}
\end{table}


\begin{figure}
\centering
\includegraphics[width=1.0\columnwidth]{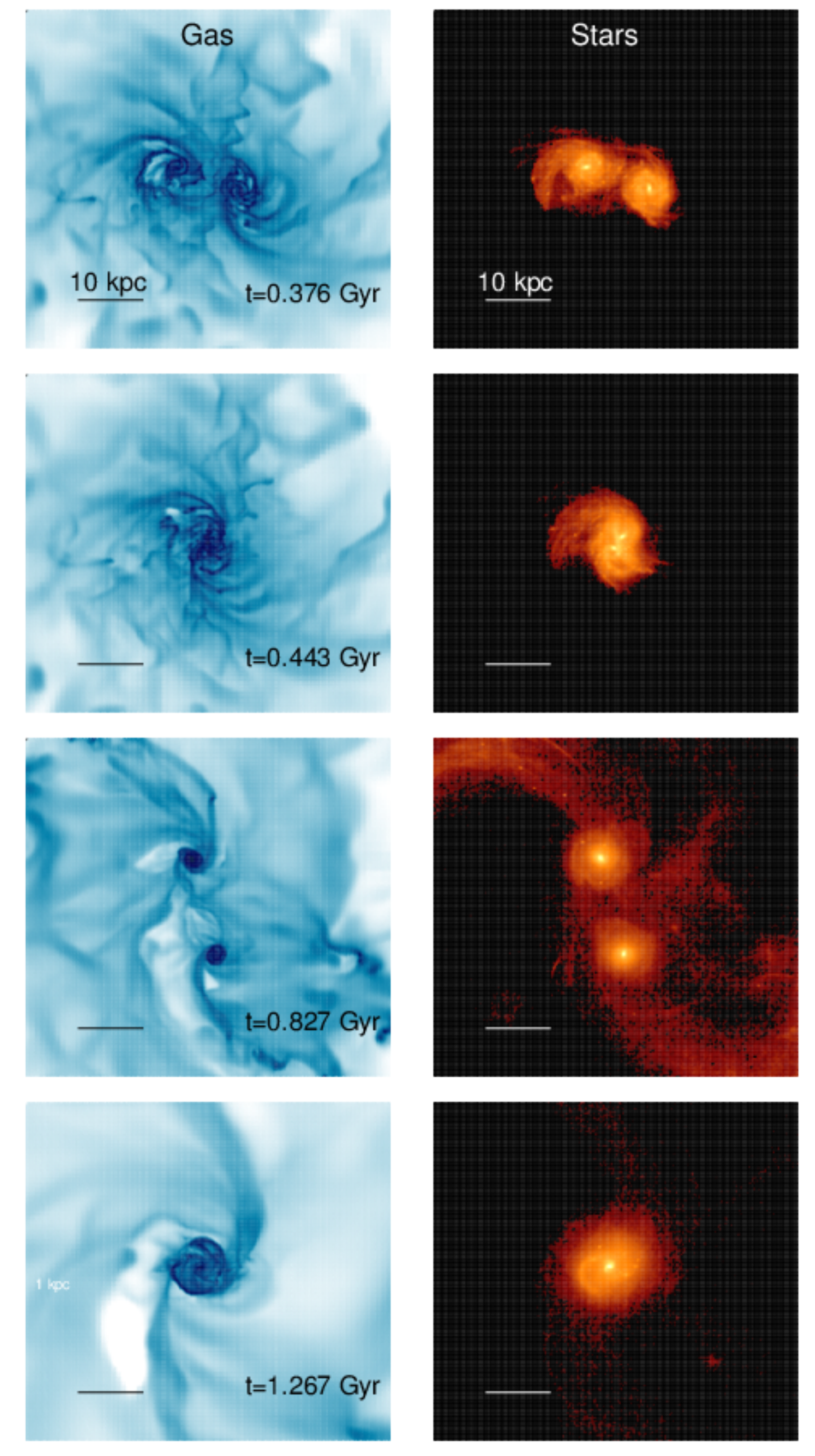}
\caption{Gas number density projection (left column) and stellar mass projection 
(right column) at different times. From top to bottom: central BHs are at 10 kpc, 
the first pericenter, the first apocenter and the point of coalescence}
\label{f1}
\end{figure}

\subsection{Gas physics}

The simulation was performed with the cosmological N-body
hydrodynamical code RAMSES \citep{Teyssier2002}. The code uses adaptive
mesh refinement, and solves the Euler equations with a second-order 
Godunov method and MUSCL scheme using a MinMod total variation diminishing 
scheme to reconstruct the cell centered values at cell interfaces.

The galaxies were set inside a computational box of ${\rm L_{box}}=400$ kpc.
The coarse level of the simulation corresponds to $\ell_{\rm min}=7$ and 
$\Delta x_{\rm coarse}=3.125$ kpc. We allowed 10 levels of refinement to get 
a maximum resolution at $\ell_{\rm max}=17$ of $\Delta x_{\rm min}=3.05$ pc.
The refinement is allowed inside a cell if i) its total mass is more than 8 
times that of the initial mass resolution, and ii) the Jeans length is resolved 
by less than 4 cells \citep{Trueloveetal1997}. If we take into account grid regions where number density is above 0.01 cm$^{-3}$, the worst cell refinement we find is at level 12 with $\Delta x \approx 97$ pc, and cells at refinement level 14 and 15 account for almost $\sim 60\%$ of the total number of cells throughout 
such areas. Above these, at level 16 we account for the $\sim 16\%$ of cells and at 
level 17 we account for $\sim 8\%$.

Our simulation includes optically thin (no self-shielding) gas cooling 
following the \citet{SutherlandDopita93} model down to temperature 
$T=10^4$ K with a contribution from metals, assuming a primordial 
composition of the various heavy elements. Below this temperature, 
the gas can cool down to $T=10$ K due to metal line cooling 
\citep{DalgarnoMcCray72}.

We adopted a star formation number density threshold of 
$n_0=250\,{\rm H\,cm^{-3}}$ with a star formation efficiency 
${\rm \epsilon_\star=0.03}$ \citep[e.g.][]{RaseraTeyssier2006,DuboisTeyssier2008}.
When a cell reaches the conditions for star formation, stellar 
(population) particles 
can be spawned following a Poisson distribution with a mass resolution 
of $m_{\star,\rm res}\approx2\times10^2$ $\rm M_\odot$. In order to 
ensure numerical stability we do not allow cells to convert more than $50\%$ 
of the gas into stars within a single time step.

After 10 Myr the most massive stars explode as SN releasing a specific 
energy of $E_{\rm SN}=10^{51}$ erg/10 M$_\odot$,
returning 20 per cent of 
the stellar particle mass back into the gas with a yield 
(fraction of metals) of $0.1$ 
inside a sphere of $r_{\rm SN}=2\Delta x_{\rm min}$. 
In order to capture the delay of stellar feedback energy release from non-thermal processes, we used 
the delayed cooling implementation of SNe feedback \citep{Teyssier+2013}. 
In this work we use $t_{\rm diss}\approx 0.25$ Myr and the energy threshold 
$e_{\rm NT}$ is the one associated to a turbulent velocity dispersion 
$\sigma_{\rm NT}\approx 50$ km/s, consistent with our resolution 
\citep[see][for details]{Dubois+2015,Prieto&Escala2016}.

\begin{figure}
\centering
\includegraphics[width=1.0\columnwidth]{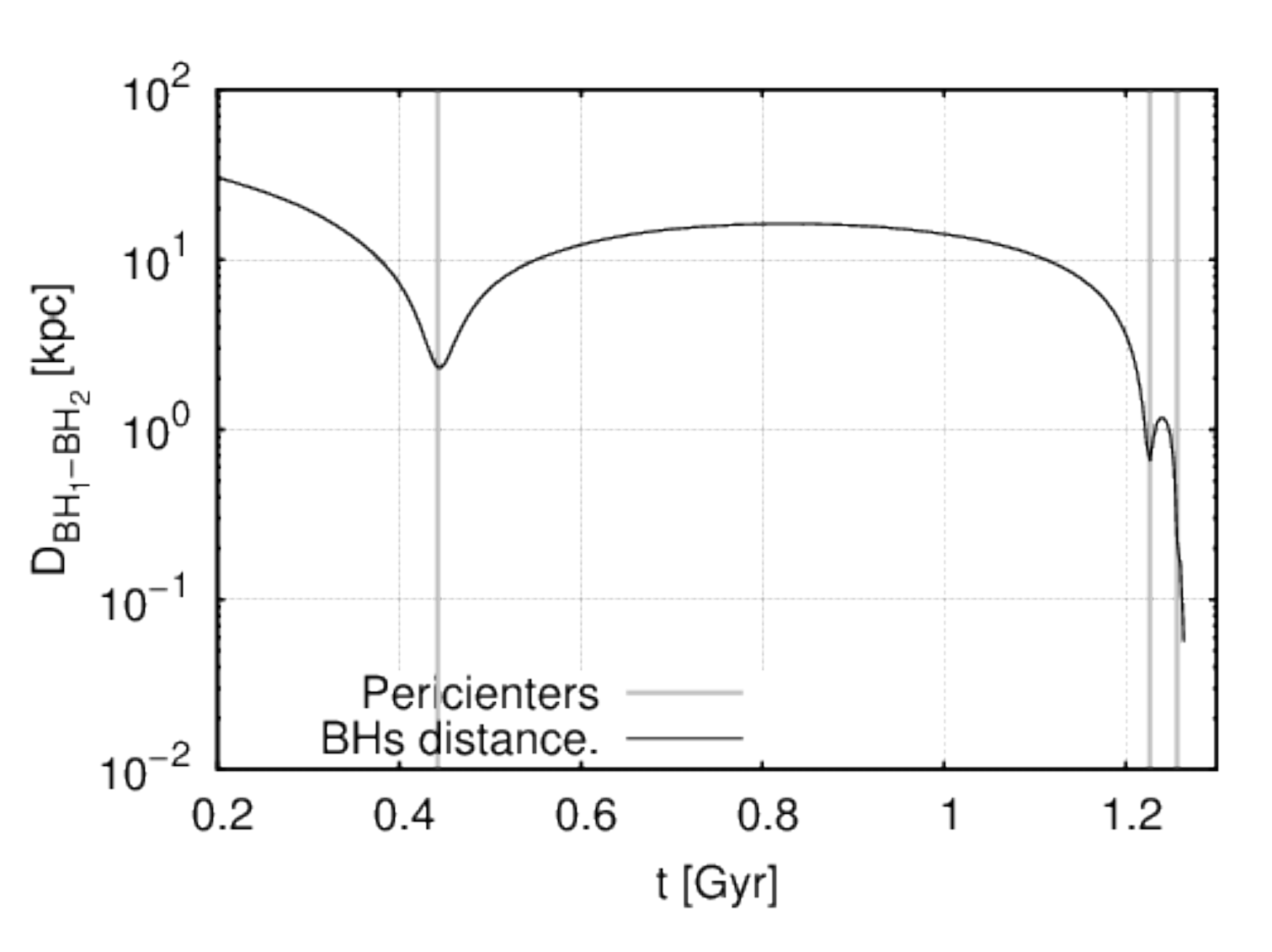}
\caption{BH particles distance as a function of time. The different gray 
vertical lines mark the first, second and third pericenter passes of the orbit 
(from left to right). After the third pericenter the BHs approach 
each other until they finally merge at ${\rm t_{merger}}=1.25$ Gyr, 
$\sim$ 800 Myr after the first pericenter pass.}
\label{f2}
\end{figure}

In order to follow the evolution of the central BH in the galaxies, 
we used sink particles \citep{Bleuler&Teyssier2014}.
We computed the mass accretion rate onto the BH using the standard
Bondi-Hoyle \citep{Bondi1952} model, $\dot{M}_{\rm Bondi}$.
In such accretion implementation the gas density is computed as an average
weighted value taken from the sink's cloud particles using a kernel following \citet{Krumholz+2004} as 
presented in \citet{Dubois+2012agn}. Throughout the simulation we cap 
the accretion rate at the Eddington rate.
The initial BH mass for both galaxies is $M_{\rm BH}=10^6\,{M_\odot}$,
approximately lying on the $M_{\rm BH}-\sigma_{\star}$ relation. Assuming a sound
speed of $c_s\approx 10-30$ km/s (for a gas temperature
$T_{gas}\approx 10^4-10^5$ K, note that most of the time the gas is
at $10^4$ K, the $10^5$ K are associated to AGN activity when the gas
is expelled from the BH vicinity), the BH influence radius at the 
beginning of the simulation is  
$R_{BH}=G\,M_{BH}/c_s^2\approx\,420-42\,{\rm pc}\,\approx\,140-14\,\Delta x_{\rm min}$.
In other words, we can resolve the BH influence radius with several cells
(note that such radius increases throughout the simulation). In addition to the forementioned grid refinement criteria, we impose that a $20$-sided cubic volume surrounding sink particles, stays fixed at maximum spatial resolution, helping to resolve the BH influence radius and to account for any potential non trivial physical processes occurring nearby.
Finally, the BHs particles 
merge if their separation is lower than $d_{\rm merge}=2\Delta x_{\rm min}$ and if they are gravitationally bound.

We have also included AGN feedback from the central BHs. AGN feedback is modeled 
with thermal energy input \citep{Teyssier+2011,Dubois+2012agn}. The rate of 
energy deposited by the BH inside the injection radius 
$r_{\rm inj}\equiv 4\Delta x_{\rm min}$ is
\begin{equation}
\dot{E}_{\rm AGN}=\epsilon_{\rm c}\epsilon_{\rm r}\dot{M}_{\rm BH}c^2.
\end{equation}
In the above expression, $\epsilon_{\rm r}=0.1$ is the radiative efficiency 
for a standard thin accretion disc \citep{SS1973} and $\epsilon_{\rm c}=0.15$ 
is the fraction of this energy coupled to the gas in order to reproduce the 
local BH-galaxy mass relation~\citep{Dubois+2012agn}. As explained in 
\citet{BoothSchaye2009}, in order to avoid  gas over-cooling the AGN 
energy is not released instantaneously every time step $\Delta t$ but 
it is accumulated until the surrounding gas temperature can be increased 
by $\Delta T_{\rm min}=10^7$ K. In order to reduce the heating effect of
the AGN we have included an extra multiplicative factor of 0.1 to $\dot{E}_{\rm AGN}$, which is done as otherwise the feedback is too effective at preventing accretion onto the central BHs, and is consistent with the scaling of radiative efficiency and BH mass found in \citet{DavisLaor}.
Such factor can be interpreted as a lower radiative efficiency, a lower 
energy coupling or a combination of both effects. This lowering of factor feedback is consistent with how NGC 2623's energetics are dominated by star formation over AGN feedback \citet{Privon+2013}.

\begin{figure}
\centering
\includegraphics[width=1.0\columnwidth]{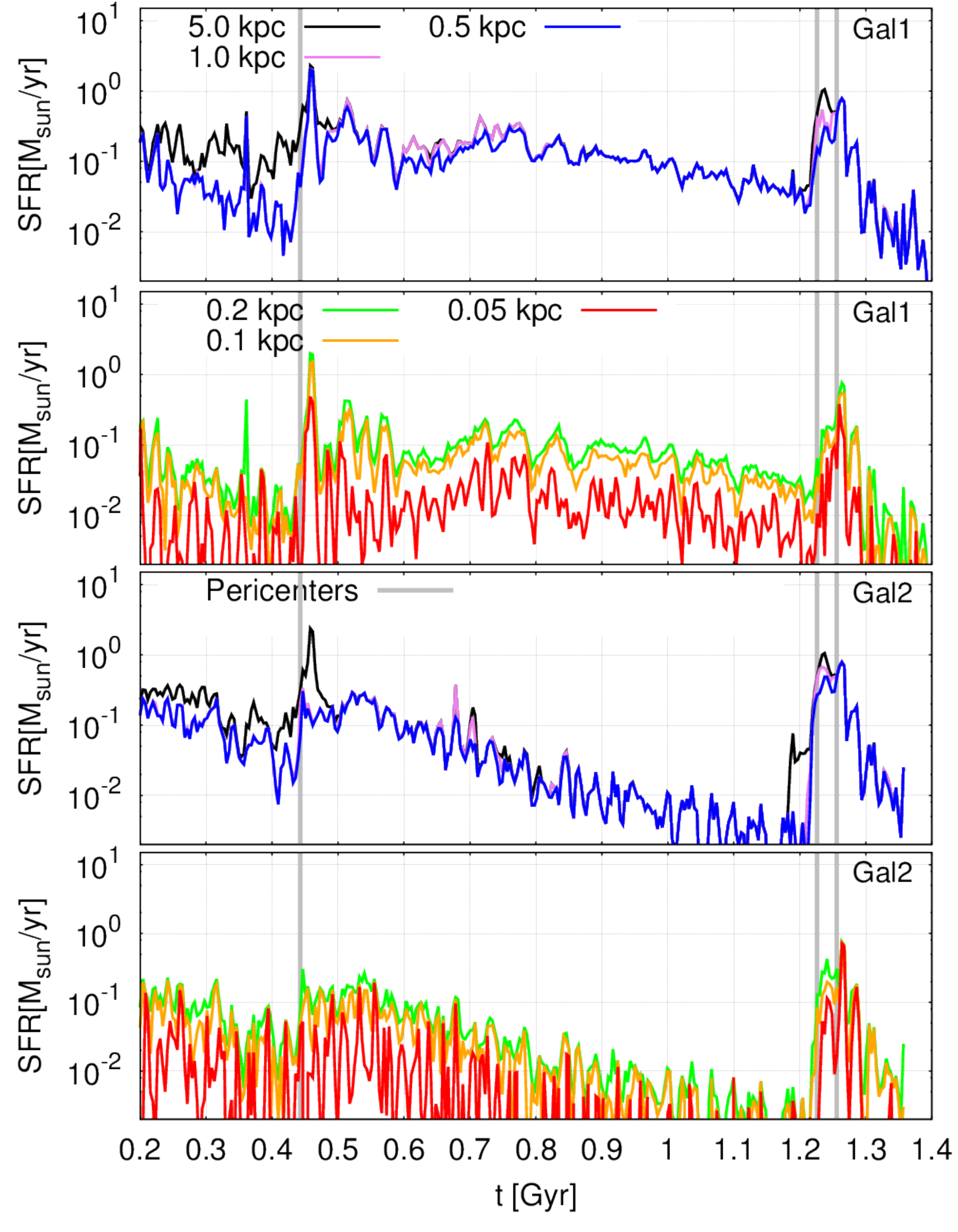}
\caption{Enclosed SFR at different distances from BH particle for both galaxies 
as a function of time: Inside 5 kpc in black, 1 kpc in violet, 0.5 kpc 
in blue, 0.2 kpc in green, 0.1 kpc in orange and 0.05 kpc in red. The gray 
solid vertical lines mark the pericenters of the orbit. It is clear that after each pericenter pass
there is a SF burst. Both the second and the third pericenter passes trigger nuclear 
SF on scales below few 100 pc whereas the first one produces more extended SF 
at $\sim$ kpc scales.}
\label{f3}
\end{figure}

\section{Results}
\label{results}

Figure \ref{f1} shows four different snapshots throughout the merger, namely (from 
top to bottom) when the BHs are at a distance of 10 kpc, the first pericenter, the 
first apocenter and time where systems coalesce (marked by the time in which the SMBHs merge). After the first pericenter 
pass the system develops two prominent tidal tails producing a ``double-tailed'' object, which can be appreciated in how the system looks like, at the point of its first apocenter in figure \ref{f1}.


Before we show results related with SF properties, BH growth and gas dynamics 
throughout the merger, it is illustrative to look at the BH separation evolution 
shown in figure \ref{f2}. This quantity is a good proxy for the galactic center
separation. The figure shows the time for pericenter passes (in solid gray vertical 
lines). After the third pericenter the BHs start to orbit around each other, 
rapidly decreasing their separation until they merge at ${\rm t_{merger}}=1.275$ Gyr. 
This time corresponds to $\sim$ 800 Myr after the first pericenter.
Furthermore, the simulated time of the BH merger depends on the minimum separation adopted (which in our case is able to resolve the sphere influence of the BHs)
for the BH coalescence; if the minimum separation is further decreased the BHs will spend more 
time orbiting each other. The pericenter passes marked by vertical lines in the figure 
\ref{f2} will guide our discussion in the following lines.

\subsection{Star formation rate}
\label{SFR}

A number of works have shown how galactic mergers-interactions trigger bursts of SF
\citep[e.g.][]{BarnesHernquist1991,Cox+2006,DiMatteo+2007,Hopkins+2008,Moreno+2019}.
The SF in mergers is not restricted to the galactic nuclear 
regions (the inner $\sim$ kpc) but it can also be triggered in gaseous 
tails of the system \citep[e.g.][]{Soifer+1984,Keel+1985,Lawrence+1989}. In 
this analysis we will focus on the nuclear (not tail) SF burst produced by 
enhancement of gas density due the galactic interactions and we will 
not study the 
extended, in-tail SF \citep[e.g.][]{Barnes2004,ChienBarnes2010,Renaud+2014}. 

Figure \ref{f3} shows the enclosed SFR inside a given radius of 
both galaxies 
throughout the evolution. The SFR is computed as the ratio between the total 
stellar mass produced in the last $\sim$ 3.75 Myr and a characteristic time defined as the 
mass weighted stellar age:
\begin{equation}
t_{\rm \star, avg}=\frac{1}{\sum_i m_{\star,i}} \sum_i t_{\star,i} m_{\star,i}.
\end{equation}
Then,
\begin{equation}
{\rm SFR}=\frac{1}{t_{\rm \star,avg}}\sum_i m_{\star,i},
\end{equation}
where $m_{\star,i}$ and $t_{\star,i}$ are the new stellar population mass and 
its age, respectively. We computed the SFR inside a sphere centered at the 
BH position for different radius, namely $R_{\rm SFR}=5,\,1,\,0.5,\,0.2,\,0.1,\,0.05$
kpc.

\begin{figure}
\centering
\includegraphics[width=1.0\columnwidth]{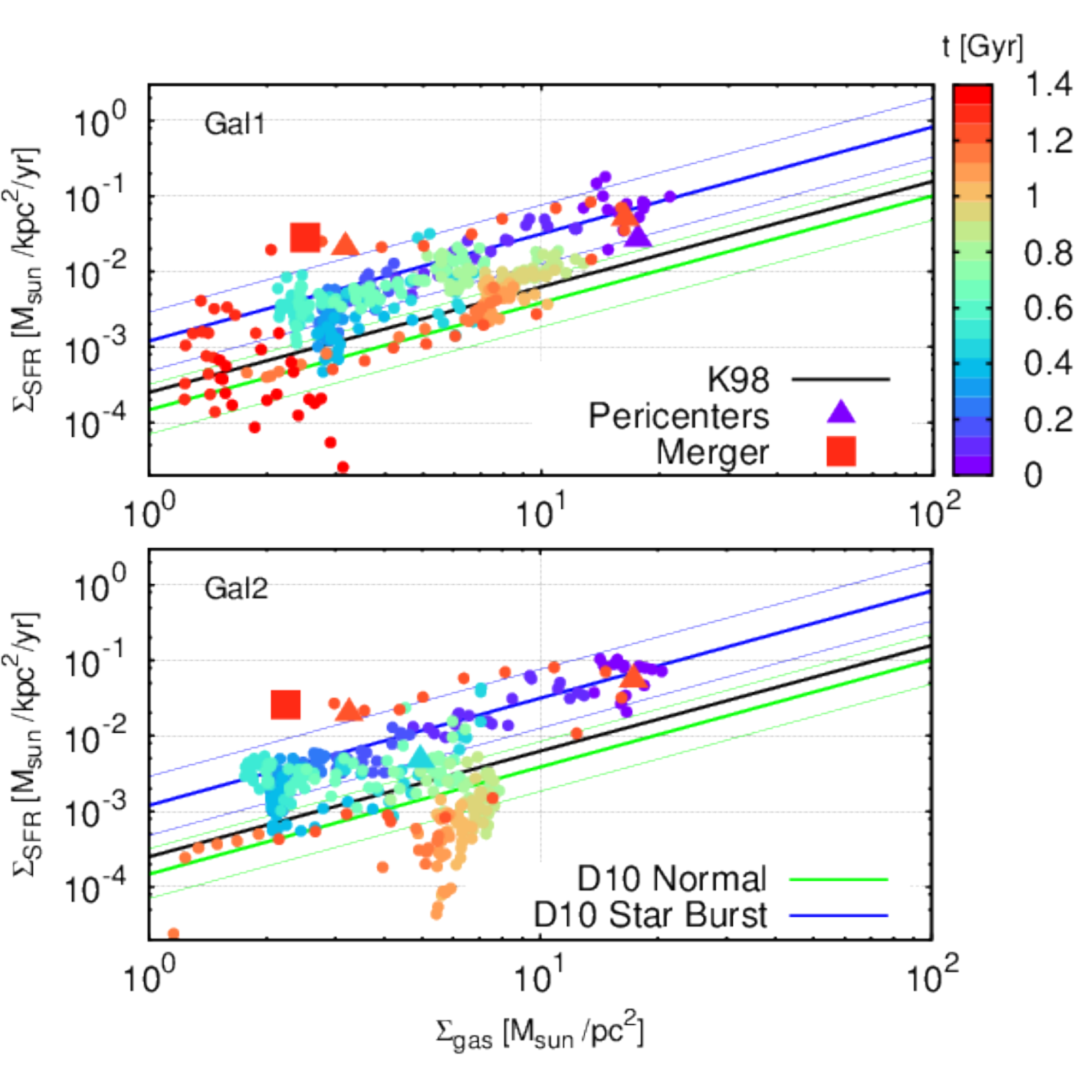}
\caption{Kennicutt-Schmidt relation for both galaxies in small circles. The 
different colors mark different times. In black solid line the 
\citet{Kennicutt1998} relation, in green solid line the \citet{Daddi+2010} 
relation for normal galaxies and in blue solid line the \citet{Daddi+2010} 
relation for star bursts. The thin lines mark the error of the corresponding
relation.}
\label{f5}
\end{figure}

The SFR in figure \ref{f3} shows intermittent behavior due to both 
stellar and BH feedback. Before any pericenter passages between the galaxies, at scales of $\sim$ few kilo-parsec, formation rates generally fluctuate 
around $\sim$ 0.1 $M_\odot/$yr. We see SF episodes lasting
 $\sim 30-40$ Myr after the pericenter passes. These relatively short periods 
are explained by the response of the medium to feedback from the massive stellar particles. We also note that star formation evolves differently for both galaxies after their first encounter. 
At $\sim$ 70 Myr before the first pericenter pass (376 Myr after the start of the simulation, see the first image of figure \ref{f1})
the spiral arms of the galaxies start to collide (at this time galactic
center distance is 10 kpc). This working interface progressively increases the
gas density, translating to a burst of SF. 

Once the galaxies reach the first pericenter a clear SF burst appears across all distance scales in Galaxy 1, with the SFR reaching $\sim$ few $M_\odot/$yr at distances over $0.1$ kpc, showing an enhancement of formation rates exceeding two orders of magnitude, which stems from the nuclear regions of the system. 
This strong burst lasts around $\ga$ 30 Myr (longer for bigger scales), after which the system is left with a relatively high but slowly decreasing nuclear SFR. In the case of Galaxy 2 there is a not too drastic increase in nuclear star formation at $0.1-0.2$ kpc after the first passage, but like for the first system, there is also a delayed strong SF episode in the outer region of the galaxy, which is itself a response to the galactic arms colliding. The collision of the spiral arms evolves more slowly than the central region collisions, which explains why the high SFRs at large scales are more persistent than the nuclear ones at this stage. After this first passage we see a steady decline in star formation for galaxy 2 (at least until another close passage happens).

The second and third pericenter encounters show prominent increases in the SF activity 
in both galaxies, where a clear increase is seen on all galactic scales. The second encounter shows a bigger SF enhancement on extended scales at first and then delayed at a close third pericenter passage, star formation from the nuclear regions becomes a highly prominent feature. These high star formation rates are maintained for around $50-100$Myr, after which, the central black holes are merged, and while the system starts stabilizing, star formation declines. 

\begin{figure}
\centering
\includegraphics[width=1.0\columnwidth]{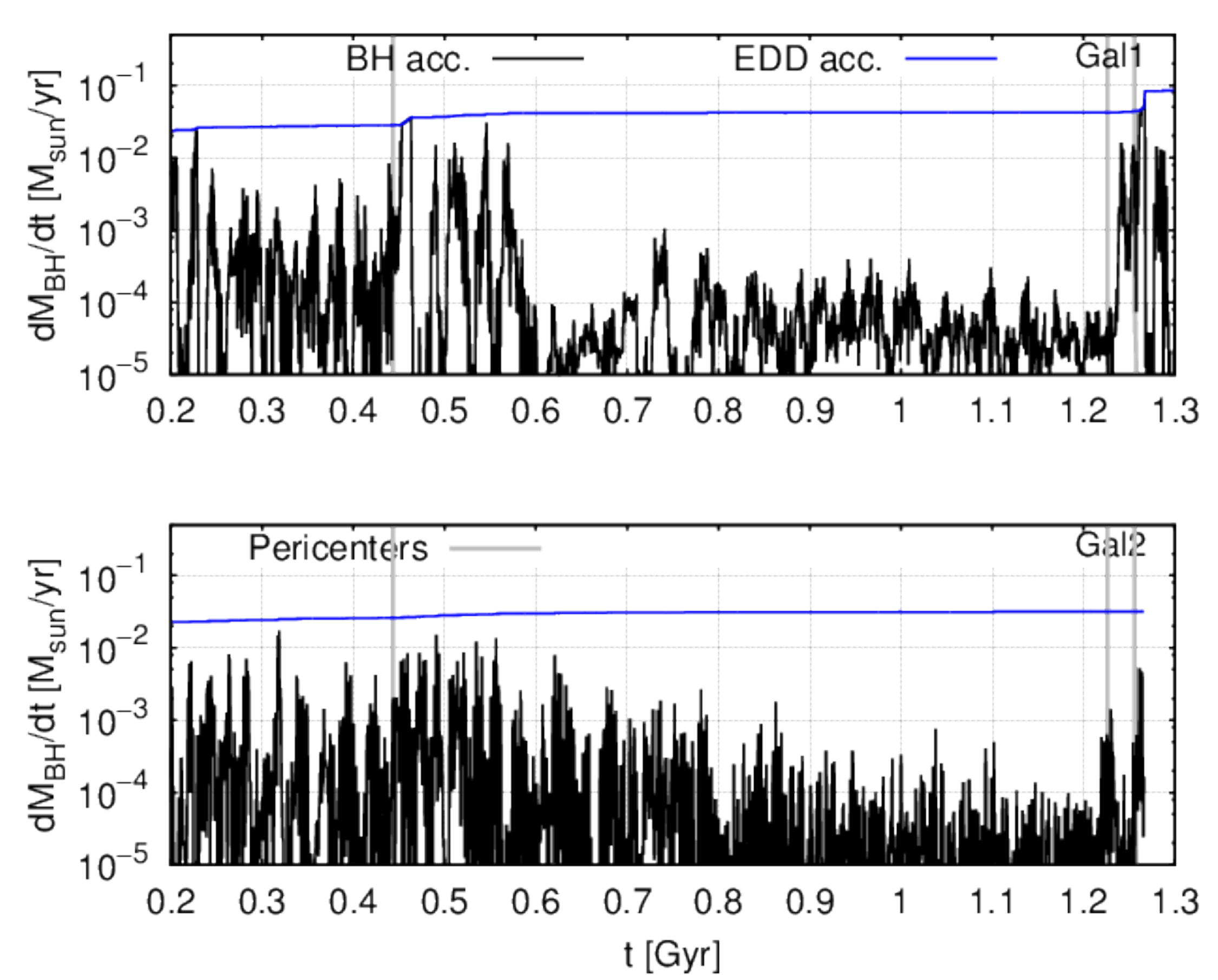}
\caption{Mass accretion rate onto the BHs in black solid 
line. The solid vertical gray lines mark the pericenters of the orbit. 
The Eddington mass accretion rate limit is in blue solid line.
Both the second and the third pericenter produce a clear increase
in the BHAR.}
\label{f6}
\end{figure}

These results show a clear correlation between pericenter passes and increases in 
SF. The SF bursts are localized to different galactic regions depending on the 
stage of the merger: the first passage trigger extended (above $\sim$ kpc, due 
to spiral arms collision) SF whereas the second and third pericenter produce 
new stars at the nuclear region (inside $\sim$ 0.5 kpc). 

 The SFR values reached in the simulation at the pericenter are 
in the range of $1-10\,M_\odot/$yr, below the SFRs measured by 
\citet{Evans+2008} corresponding to $\sim\,50-90\,M_\odot/$yr, and also below 
the $\sim\,70\,M_\odot/$yr found by \citet{Howell+2010} for a system with similar initial conditions as ours (NGC 2623).
The simulated values are closer to 
the  $8\,M_\odot/$yr rate found for the system's recent past in 
in \citet{Cortijo+2017}. This SFRs values are realistic for a merger system \citep{Pearson+2019}, albeit it would put our simulation below typical starburst galaxy rates.


Figure \ref{f5} shows the Kennicutt-Schmidt relation
\citep[][here after K98]{Schmidt1959,Kennicutt1998} for both galaxies as 
a function of time. In order to compute the galactic disc surface density 
$\Sigma_{\rm gas}$ and the surface SFR $\Sigma_{\rm SFR}$ we have defined 
a radius $R_{\rm disc}$ and a height $h_{\rm disc}$ inside a box with 12 kpc 
of side centered at the BH position; the SFR is computed within this cylinder.
The equatorial plane of the cylinder 
is constructed with a point and a normal vector, namely the BH position and 
the gas angular momentum vector computed inside 2 kpc from the BH position
(see appendix \ref{appB} for a discussion about rotational center). Inside 
the 12 kpc side box we compute the enclosed mass in both the positive and 
the negative $\hat{z}$ direction as a function of height $z$. The disc height 
$h_{\rm disc}$ corresponds to the altitude $z$ where the cylinder contains 
90\% of the baryonic mass. Following an analogous method we computed the radius 
$R_{\rm disc}$ as the radius where for a cylinder height $h_{\rm disc}$ the 
disc contains 90\% of the baryonic mass. After this procedure we define
\begin{equation}
\Sigma_{\rm gas}=\frac{M_{\rm gas}}{\pi R_{\rm disc}^2},
\end{equation}
where $M_{\rm gas}$ is the gas mass inside the cylinder and 
\begin{equation}
\Sigma_{\rm SFR}=\frac{{\rm SFR}}{\pi R_{\rm disc}^2},
\end{equation}
with the SFR computed each $\sim$ 3.75 Myr.

Both systems start near the \citet{Daddi+2010} 
(hereafter D10) Star Burst SF relation and evolves between this and the usual K98 relation (blue-cyan dots) showing no clear transition behaviour around the first merger passage. After around $\sim\,600$ Myr (green to orange transition) where we found a steady decline in SFRs at figure \ref{f3}, the galaxy 2 system starts going below both the K98 and D10 relations, exhibiting how star formation is not able to keep up with the amount of gas stripped from the galaxy by the merger interaction. 
Later pericenter passages bring both galaxies above the star burst D10 relation, and after the violent episodes of both SF and AGN feedback that the systems are subjected to,
the eventually merged galaxy evolves progressively to the region below the D10 normal galaxies, showing a clear decrease in SFR \citep[see][]{Renaud+2014}.

\begin{figure}
\centering
\includegraphics[width=1.0\columnwidth]{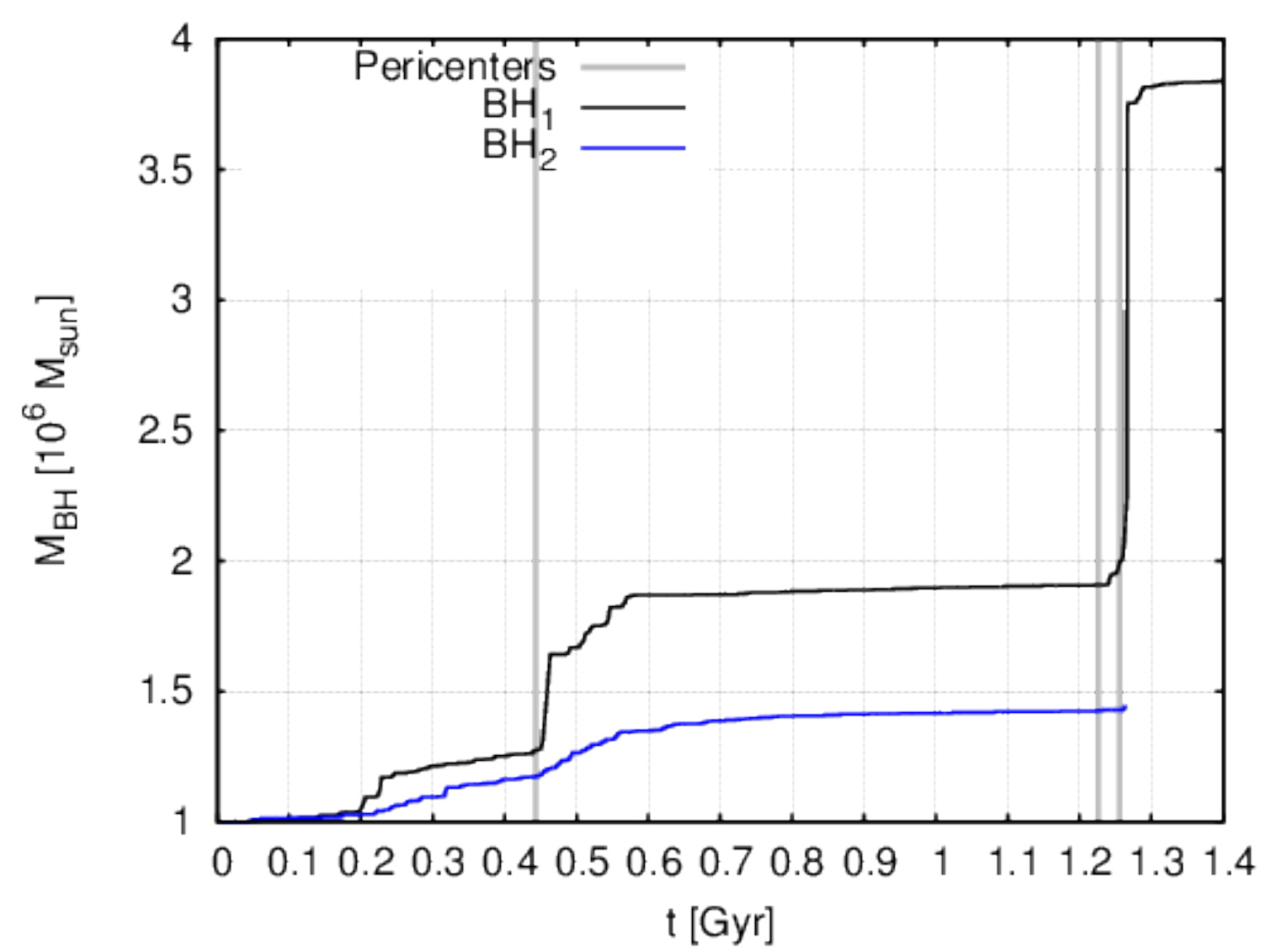}
\caption{BH mass evolution. In solid black line the BH mass associated to 
galaxy 1 and in solid blue line the BH mass associated to galaxy 2. The 
solid vertical gray lines mark the pericenters of the orbit.}
\label{f7}
\end{figure}

\subsection{Black hole evolution}

Observational evidence suggests that galactic encounters 
can trigger AGN activity \citep[e.g.][]{Veilleux+2002,Giavalisco+2004,Treister+2012}. 
In order to feed the BHs the galactic gas should reach the sphere of influence of 
the central massive objects. To accrete on to the BH, gas orbiting around 
the BH must lose angular momentum, resulting in an inward gas mass flow in the galaxy.
This can be triggered by gravitational torques acting on the gas due to the galaxy-galaxy
interaction \citep[e.g.][]{Barnes1988,BarnesHernquist1991,DiMatteo+2005,Cox+2008}
or in general by any type of torque (gravitational, pressure gradients/
hydrodynamic, magnetic, or viscous) capable of changing the angular momentum of the
gaseous component of the galaxy.

Figure \ref{f6} shows the BH mass accretion rate as a function of time for both
galaxies. The black hole accretion rates (BHAR) oscillate in the range of $\sim$ $10^{-2}-10^{-4}\,M_\odot/$yr over the first $\sim600$ Myr, where we see galaxy 1 approaching (and eventually peaking at) the Eddington limit on different occasions. After the first pericenter passage, AGN feedback strongly regulates accretion rates, lowering them for at least an order of magnitude.  

\begin{figure}
\centering
\includegraphics[width=1.0\columnwidth]{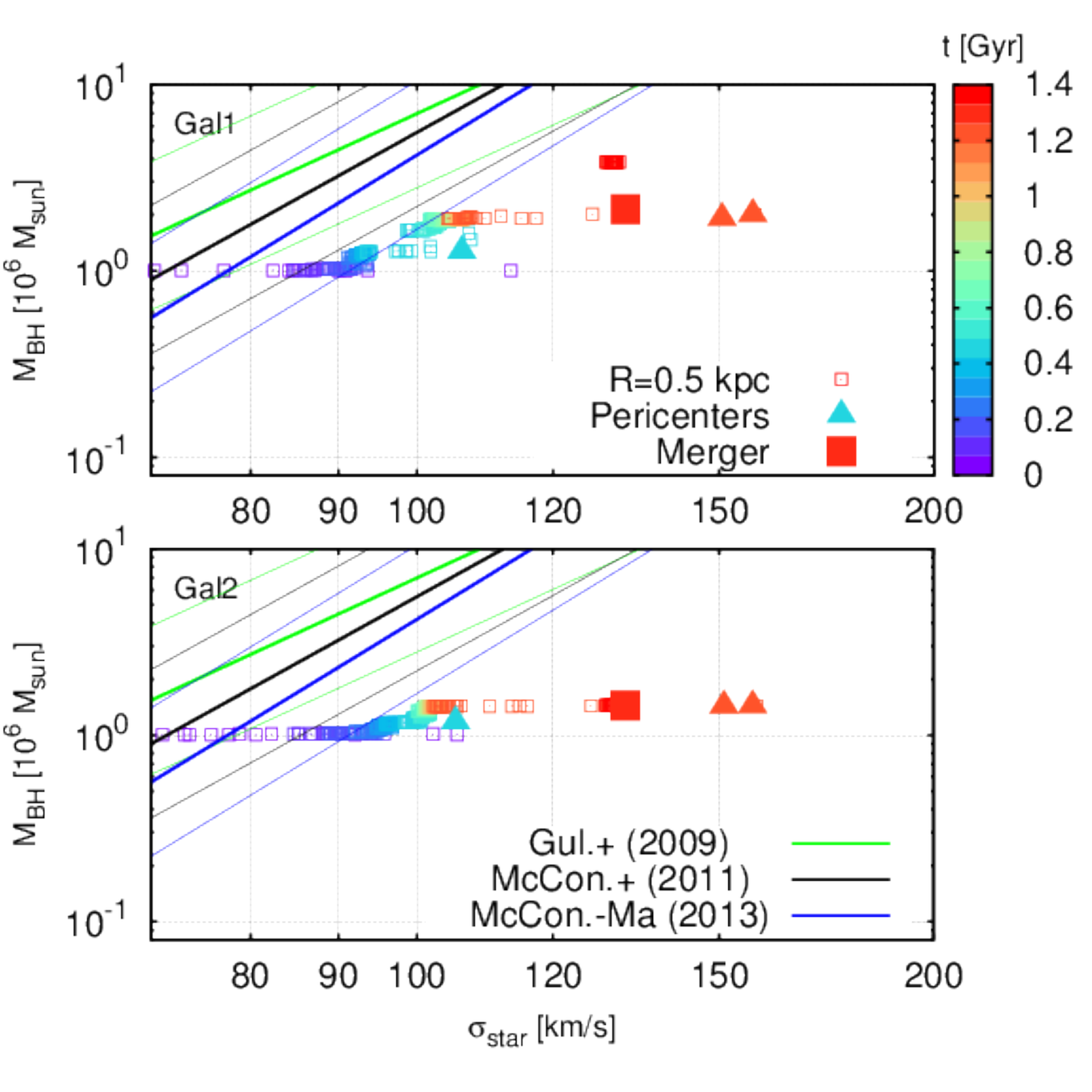}
\caption{BH mass-bulge stellar velocity dispersion relation. Different 
colors mark different times. Filled circles mark the relation taking 
into account all stars inside 1 kpc around the central BH and empty 
squares mark the relation for stars inside 0.5 kpc around the central BH.
The broad black solid line shows the \citet{McConnell+2011} relation, the
broad blue solid line shows the \citet{McConnellMa2013} relation and
the broad green solid line shows the \citet{Gultekin+2009} relation. The 
thin green, blue and black line are the corresponding relation 0.4 dex 
above and below the central one.}
\label{f8}
\end{figure}

Following the low BHAR after the first pericenter passage, in the two following passes, both systems exhibit clear peaks due to the funnelling of gas towards the central galaxy regions. These peaks are more pronounced on galaxy 1 than in galaxy 2, the first reaching BHAR values of a few $10^{-2}\,M_\odot/$yr, whilst the second only has low (albeit pronounced) peaks of a $10^{-3}\,M_\odot/$yr. After these two last encounters the 
BHs merge (where just before this, galaxy 1 accreted at the Eddington rate for a short period of time).  

Figure \ref{f7} shows the BH masses as a function of time. Because 
the differences in their mass accretion rate the BH masses are also different 
for both objects. The BH$_1$ mass shows a clearer increase with the 
first pericenter compared with BH$_2$, as can be seen from its mass accretion
rate (figure \ref{f5}). We confirm that after undergoing a strong mass gaining episode, BH$_1$'s mass stays nearly constant for $\sim 600$Myr, with a mass of $\sim1.8\times 10^6\,M_\odot$. During this 
time interval the galaxies have reached their first pericenter producing an 
enhancement in the BH$_1$ mass accretion rate and consequently in its mass.
This BH growth is not associated with galactic bulge coalescence and show that
BHs can grow in stages before the galactic bulge merge \citep{Medling+2015}.

\begin{figure}
\centering
\includegraphics[width=1.0\columnwidth]{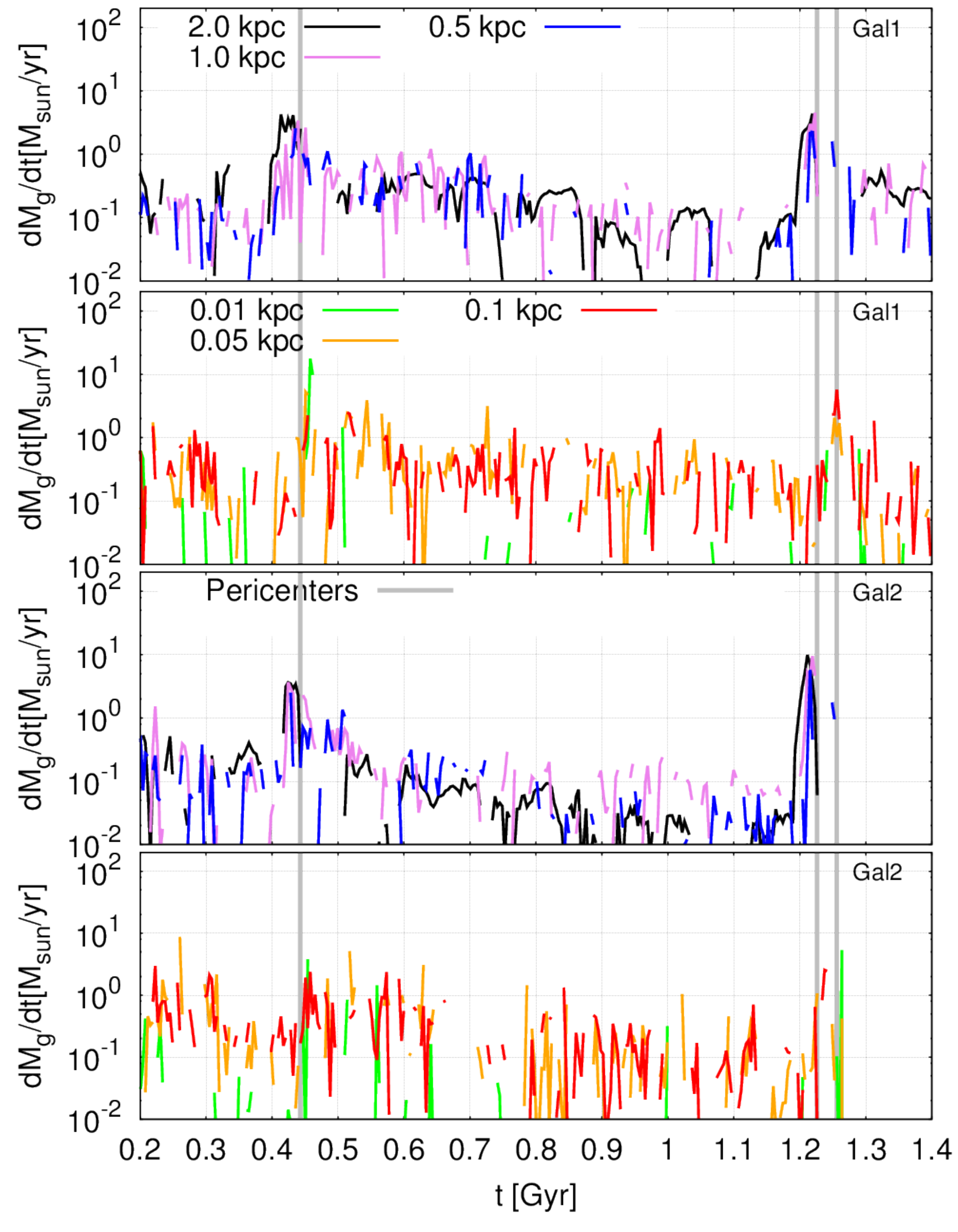}
\caption{Inward gas mass accretion rate as a function of time at different
distances from the galactic stellar center of mass: 2 kpc in black, 1 kpc 
in violet, 0.5 kpc in blue, 0.1 kpc in green, 0.05 kpc in orange and 0.01 
kpc in red. The solid vertical gray lines mark the pericenters of the 
galactic orbit. The figure 
shows that pericenters correlate with peaks of mass accretion rate.}
\label{f9}
\end{figure}

In contrast to the BH$_1$ evolution, the second compact object does show 
a clear increase in the first pericenter but it is substancially lower, as can be seen from the 
low mass accretion rate shown in figure \ref{f5} (which caps at $\sim 10^{-2}\,M_\odot/$yr, which is not necessarily low, but there is a big amount of perceivable variation in rates).
The evolution becomes nearly flat, with a low amount of growth until the merger. At the time of coalescence, BH$_1$ had grown nearly twice as much as what BH$_2$ had grown through accretion, and after merger, the remnant BH ends up at $\sim 3.8\times10^6\,M_\odot$. This final value would put the final BH mass well below the LIRG galaxy mergers (like NGC 2623) found in the GOALS sample (\citet{Haan+2011}), and although initially this is neither an indication that the black holes are not accreting enough gas through the merger evolution, nor that the BH initial masses are wrong, we can further the analysis by checking how the M-$\sigma$ relation evolves throughout the simulation.

Given the BH mass at each point and the stellar velocities, it is possible 
to compute the $M_{\rm BH}-\sigma_\star$ (``M-sigma'') relation.
\citep{McConnell+2011,McConnellMa2013,Gultekin+2009}
Figure \ref{f8} shows such relation as a function of time. We have initialized 
the simulation with a BH mass $M_{\rm BH}=10^6$ M$_\odot$ and a bulge
velocity dispersion $\sigma_b\approx 110$ km$/$s, which means our setup is inside the empirical relation of \citet{McConnellMa2013} when using the velocity dispersion from the $<0.5$kpc region. Even though it is normal for velocity dispersion to grow quickly in proportion to BH mass in a merger process due to the strong dynamical perturbation the bulges suffer (and therefore we expect a tendency that the measured M$-\sigma$ values should partially stray to the right of the relation), we see in the figure both galaxies quickly moving far away from the empirical relation. This means that the feeding of the BHs is not being able to catch up with the growth of velocity dispersion (it should be noted that after the stellar bulges merge, velocity dispersion should not increase, and BH feeding could slowly bring the system back to the empirical relation as the galaxy stabilizes).

To further support how BHs are growing less than what is expected of them, we see for instance, that BH$_1$ grows from $10^6\,M_\odot$ to $\sim 2\times 10^6\,M_\odot$ in $1.2$ Gyr, and if the Eddington accretion rate is $\dot{M}_{\text{Edd}}(t)=\frac{M_{\text{BH}}(t)}{t_{\text{Sal}}}$ (with the Salpeter timescale being $t_{\text{Sal}}=4.5$ Myr for our radiative efficiency), we would have an average accretion rate of $\approx 2,5\times 10^{-3}\dot{M}_{\text{Edd}}$, well under typical feeding expected for radiative AGN feedback to be relevant. 

The apparent culprit of these overall lower than expected average BHAR, would be the amount of thermal feedback being put back into the grid, which heats the gas surrounding the vicinity of our BHs too effectively for accretion to be steadily maintained. This is further evidenced by how even though torques at the hill radius are sustained all through the simulation (see section 3.4), this does not translate into a feeding of the black holes, as we see instead that the only important feeding episodes occur in the initial stages of the simulation and at close passages (where material is too efficiently transported towards the center, allowing gas dynamics to overcome the heating effect of feedback). 

The straightforward AGN feedback approach that we are using from \citet{Dubois+2012agn} was developed for cosmological simulations, and even though it has seen successful use in that context, the main difference here is that at the high resolutions we achieve, such simple recipe may result in the failure to capture the correct small scale physics that model the heating of the central bulges by the BHs. It has been shown that different methods for dealing with BH feedback may yield quite different results, and that direct injection of thermal energy to galactic cores may produce strong and persistent outflows or cavities in central regions that suppress accretion \citep{WursterThacker2013}. It is then imperative that we try to capture the more detailed heating structure that is produced by the radiative transfer of the soft X-ray photons that produce quasar-mode feedback. There have been successful efforts at capturing the heating rate from the expected X-ray emission of the central AGN from \citet{Choi+2012}, but this recipe is still at its heart a direct injection of energy back to the galactic core, and does not offer any accounting of radiative transfer effects.

A more consistent option for improving our feedback recipe, would be to include radiation coupling to our hydrodynamics through RAMSES-RT, in the code presented by \citet{Rosdahl+2013} and \citet{Rosdahl+2015}. Quasar feedback has already been modelled in this way \citep{Bieri+2017},  and it relies in the coupling of hydrodynamics with the radiative transfer of photons being introduced by the sink particle into the grid through an AGN template spectral distribution, allowing for a detailed accounting of the production and reprocessing of X-ray radiation (and therefore the overall heating mechanisms) produced by the innermost regions. The introduction of the RT module would also allow for a more consistent modelling of SN feedback, and presents the opportunity for future work.

\begin{figure*}
\centering
\includegraphics[width=1.0\columnwidth]{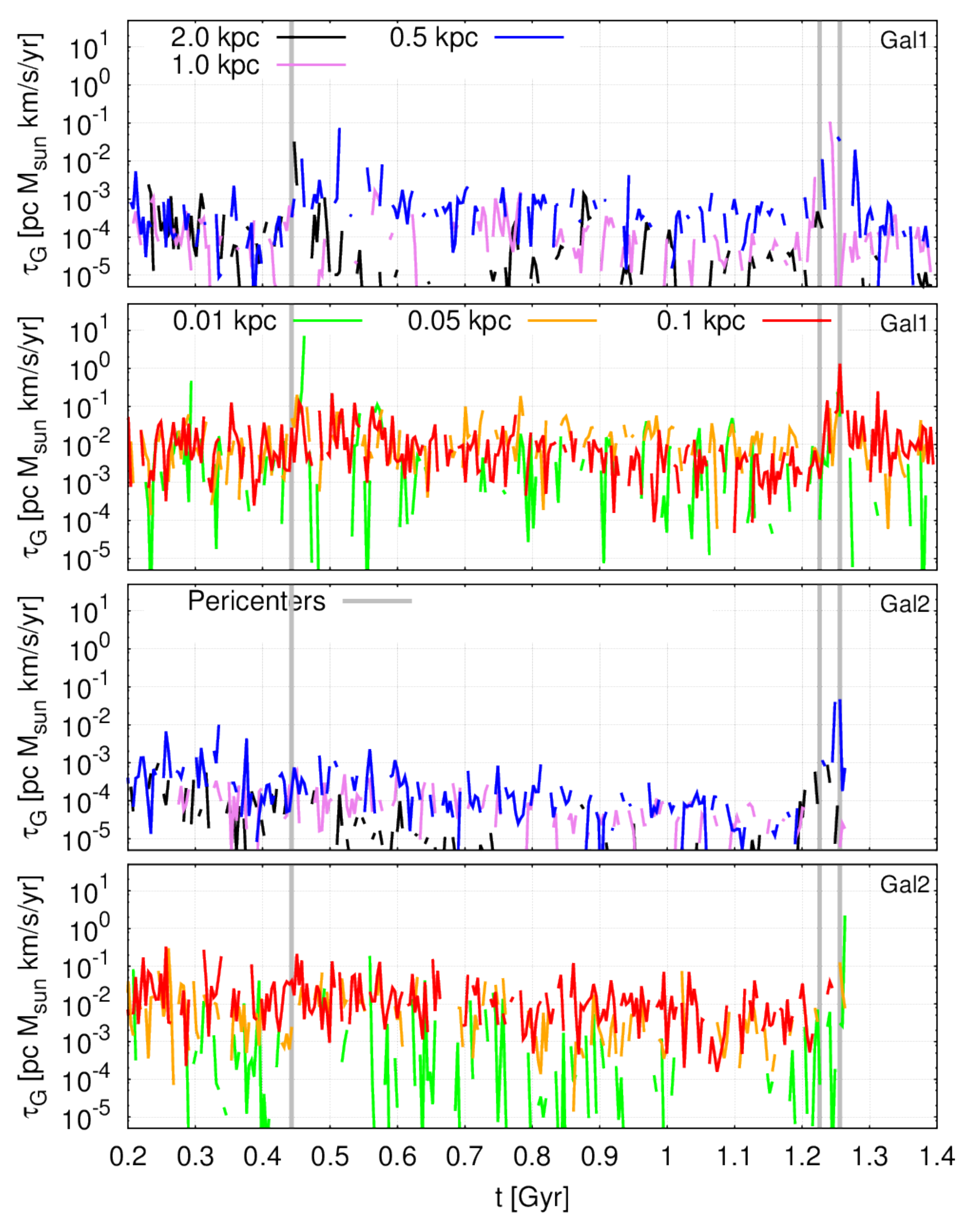}
\includegraphics[width=1.0\columnwidth]{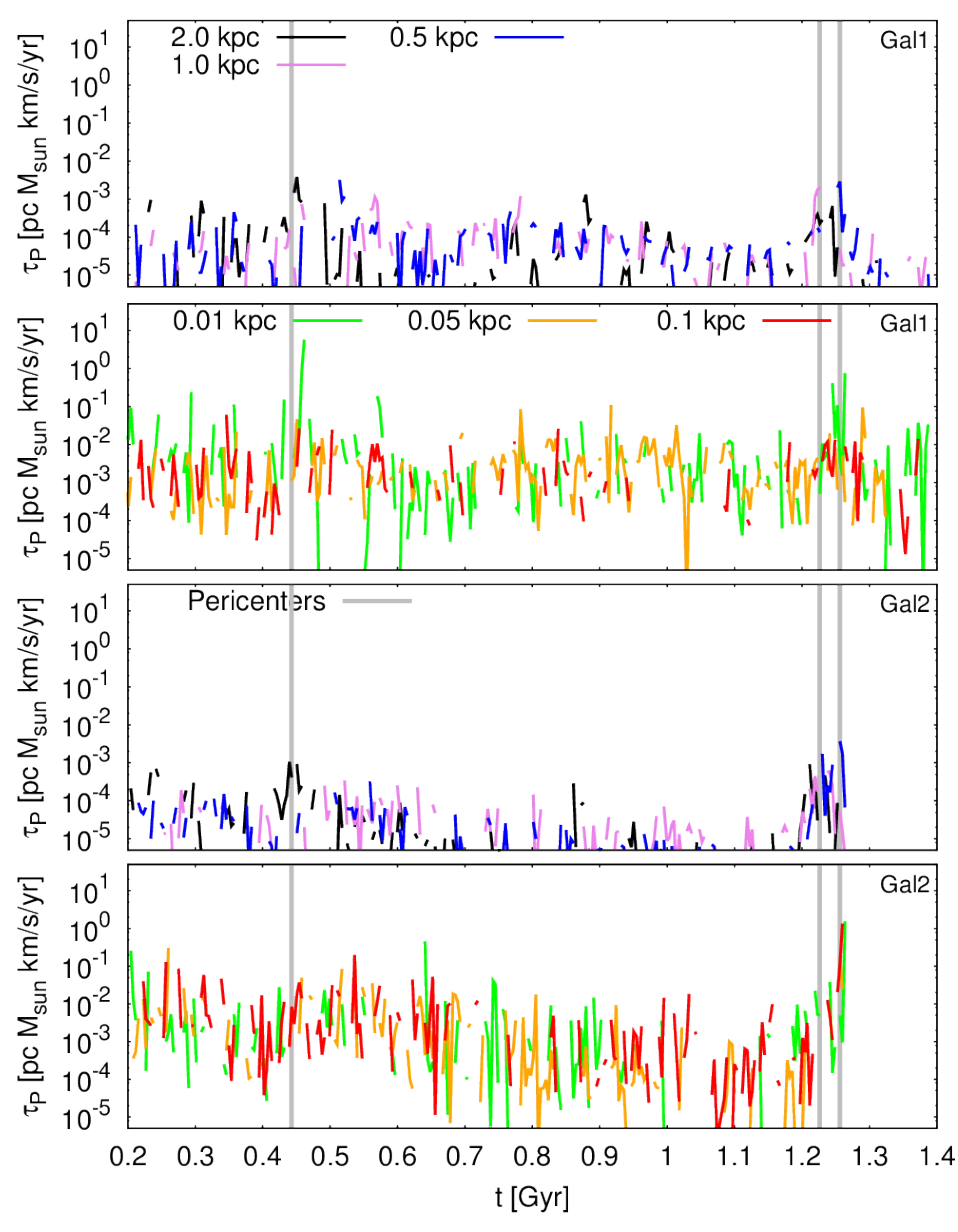}
\caption{Left: Total gravitational torque on the gas associated to inward 
mass transport at different distances from the stellar center of mass: 2 
kpc in black, 1 kpc in violet, 0.5 kpc in blue, 0.1 kpc in red, 0.05 kpc 
in orange and 0.01 kpc in green. The solid vertical gray lines mark the 
pericenters of the orbit. Right: Same as left column but for the hydrodynamic
torque. The figure shows that pericenters are associated with increases in 
torques.}
\label{f10}
\end{figure*}

\subsection{Gas accretion rate}

In the last section we showed that peaks of BH mass accretion rate 
correlate with the pericentric passages, suggesting a connection between 
close encounters and enhancement of gas inflows in galaxies. Under this scenario 
it is useful to look at the inward gas mass accretion rate at different radii. 
Figure \ref{f9} shows the inward gas mass accretion rate at different distances 
from the stellar center of mass. As with the KS computation,
we have constructed a disc perpendicular to the gas angular momentum vector.
After that, in order to compute the gas accretion rate we look for the 
stellar center of mass inside 2 kpc around the BH for each galaxy. Given the 
position $\vec{r}_{\rm CM}$ and bulk velocity $\vec{v}_{\rm CM}$ of the center 
of mass we have computed the inflowing gas mass accretion rate as
\begin{equation}
\dot{M}_{\rm g}=\sum_i \rho_i\left(\vec{v}_i-\vec{v}_{\rm CM}\right)\cdot\Delta \vec{A}_i.
\end{equation}
where $\vec{v}_i$ is the gas cell velocity and $\vec{A}_i$ is the surface 
element crossed by the gas in a direction parallel to the radial vector 
$\vec{r}_i-\vec{r}_{\rm CM}$, with $\vec{r}_i$ the gas cell position. The sum is 
computed inside an annulus of width $\Delta x_{\rm min}$ for $r\le$ 200 
pc, corresponding to the level of refinement 17 and $32\,\Delta x_{\rm min}$ for 
$r\ge$ 500 pc, corresponding to the refinement level 12. 

Figure \ref{f9} shows a clear correlation between peaks of gas 
mass accretion rate on scales $\gtrsim0.5$ kpc and close passes for both galaxies
(first and third panel from top). The first pericenter pass is associated 
with a gas mass accretion rate as high as $\sim$ $5\,M_\odot/$yr. A few Myr 
before the second pericenter pass mass accretion rates 
reach $\sim$ few $5-10\,M_\odot/$yr at large scales. Such episodes of inflowing 
mass on large scales are consistent with enhancement of SF in pericenters as shown 
in figure \ref{f3}. The causal relation between these two phenomena can be seen by
comparing the SFR and the mass accretion rate: the close passes produce mass inflows
which are followed after a few Myr by bursts of SF.

At small scales (below $\sim\,100$ pc), the enhancement in mass accretion 
rate at the first pericenter is not as significant as it is in large scales, except for very short bursts of inflow at $0.05-0.01$kpc scales in Galaxy 1. The inflowing 
mass accretion rate reaches values that are above $\sim 10 M_\odot/$yr in galaxy 1 in the nucleus at $10$ pc in this brief burst (which happens at distances below the order of the BH sphere of influence), but rates are generally around the $\sim 1M_\odot/$yr value, and are sustained in a somewhat irregular fashion before the first encounter. Galaxy 2 shows a slightly 
more consistent mass accretion rate in the same period of time at similar scales, but rates are not perceivably higher. The second and third pericenter passes show an enhancement in mass accretion at small scales. the 
amount of inflowing mass is able to trigger (after few Myr) SF bursts 
and feed the BHs as has been shown in the previous sections. In 
particular, at the third pericenter pass the gas inflow
rate approaches $\sim 10\,M_\odot/$yr due to the gas bulges coalescence.

We conclude that there is a correlation between peaks of gas mass 
accretion rate, SFR, and BHAR associated with pericenter passages. In other words, 
close galactic encounters trigger mass inflows crossing the BH influence 
radius, producing SF bursts and lighting up AGN activity in galactic centers.

\subsection{Torques on the gas}

At this point we have shown that throughout the merger process there 
are episodes of efficient gas inflows toward the galaxy centers. 
In order to fully understand the origin of mass 
transport into the galactic center it is necessary to quantify the torques 
acting on the gas, in order to link mass inflow episodes with angular 
momentum loses (see appendix \ref{appD}).

Figure \ref{f10} shows the torques acting on the galactic 
disc at different radius as a function of time. Before computing the 
torques, we have defined the galactic disc in the same way we did it to 
compute the KS relation and the gas mass transport.

We have computed the torques with respect the stellar center of mass
$\vec{r}_{\rm CM}$ as a proxy for the rotational center of each spiral galaxy 
(see appendix \ref{appB} for a discussion about rotational centers). In order 
to do that, it is necessary to set a non-rotating coordinate system free 
falling with the stars. In such a frame, the acceleration of a particle 
becomes $\vec{a}'_i=\vec{a}_i-\vec{a}_{\rm CM}$, where $\vec{a}_i$ is the 
particle acceleration with respect an inertial reference frame (the center 
of the fixed simulation box in our case) and $\vec{a}_{\rm CM}$ is the acceleration 
of the stellar center of mass with respect the same inertial frame (see 
appendix \ref{appC}). Then, in the co-moving reference frame the torques can 
be computed as
\begin{equation}
\vec{\tau}'=\sum_i m_i(\vec{r}_i-\vec{r}_{\rm CM})\times(\vec{a}_i-\vec{a}_{\rm CM}).
\label{torque}
\end{equation}
In the previous expression $\vec{r}_i$ is the cell position, $m_i$ is the 
gas cell mass. The acceleration $\vec{a}_i$ is the combination of the 
gravitational acceleration $-\nabla\phi_i$ and the acceleration associated 
to hydrodynamic on the gas $\nabla P_i/\rho_i$, where $\phi_i$ is the 
gravitational potential at a given cell and $P_i$ is the pressure in the 
same cell.

Because the galactic disc is defined in terms of the disc angular 
momentum, negative torques imply a loss of angular momentum and a 
resulting inward mass transport. Figure \ref{f10} shows $-\vec{\tau}'$, 
i.e. torques producing net inward mass transport inside an annulus at a 
given distance from the galactic centers (the regions without data are 
dominated by outward mass transport torques). The sum is computed inside 
an annulus of width $\Delta x_{\rm min}$ for $r\le$ 200 pc and 
$32\,\Delta x_{\rm min}$ for $r\ge$ 500 pc, as in the gas mass
accretion rate computation.

The left column of figure \ref{f10} shows that at large scales 
($\ga\,0.5$ kpc) both galaxies show large fluctuations in gravitational 
torques with galaxy 1 reaching larger torque values with higher 
fluctuation. In both galaxies gravitational torques are more 
important at 0.5 kpc than at larger radii. It can be understood in terms 
of the higher gas concentration at lower galactic radii.  

The huge fluctuations in gravitational torques at large scales makes difficult 
to identify a peak associated with the first pericenter in both galaxies.
In galaxy 1 it is possible to identify a coherent increase at 2 kpc about 
$\sim$ few Myr after the first passage. On the other hand, 
in galaxy 2 no gravitational torque peak can be identified at large scales. 

In galaxy 2 the late pericenter passages are associated with increased gravitational 
torques on larger scales (mainly at 0.5 kpc scales for the second passage). In galaxy 1 it is more difficult to recognize a torque increase (only having measured one relevant torque spike at 1 kpc between passages). We also see some torque presence at 0.5 kpc after the merger of the systems.

Gravitational torques acting on the galactic central region, i.e. less than 
100 pc from the stellar center of mass, show a clear enhancement associated 
with the second and third pericenter passes but an almost imperceptible change during the
first pericenter pass for both galaxies. Figure \ref{f10} shows that the inner 
galactic region, besides a very short increase of torque at 10 pc in galaxy 1 after the first passage, feels the maximum gravitational torque around the third pericenter pass (with a very big spike at the smallest scales for galaxy 2 when the systems are about to merge). Such strong torques acting on the galactic gas 
produces gas inflows and feeds the central massive objects, lighting up 
the AGN. It is also of note, that gravitational torques feature most importantly, at 100 pc scales, which aligns with how the BH influence sphere helps with gas transport at this distance.

The right column of figure \ref{f10} shows the hydrodynamic torques associated with 
inward mass transport. At large spatial scales hydrodynamic torques 
are lower and more sporadic than gravitational torques. This hydrodynamical torque values become closer to gravitational ones at pericenter passages, where especially for galaxy 2, we se features at every passage. 

Within the galactic nuclei the hydrodynamic torques match quite well with gravitational torques at the smallest scales, showing peaks in the same places where their counterparts do. Such enhancements are at the 
same level as the gravitational torques showing that both mechanisms are 
working to redistribute matter in the later stages of the merger. In 
other words, hydrodynamic torques work in tandem with gravitational torque in
order to redistribute mass and angular momentum in the galactic disc.

\section{Discussion and Conclusions}
\label{conclusion}

With the aim of studying the connection between torques and mass transport
in galactic discs, we have simulated a galaxy merger employing realistic initial conditions based in \citet{Privon+2013}. The SFR reaches values of $\sim\,1-10\,M_\odot/$yr,
below the observational measurements from \citep{Evans+2008,Howell+2010} for NGC 2623 specifically, but closer to the values presented in \citet{Cortijo+2017}, this puts our system below the star forming capabilities of a starbursting system, but inside expected rates for generic merger systems \citep{Pearson+2019}. The final BH mass of the 
system is $M_{\rm BH}\approx3.8\times10^6\,M_\odot$, around one or two order of magnitudes below the usual values presented in \citet{Haan+2011} and below  the 
dispersion of the ``M-sigma'' relation \citep{Gultekin+2009}. This low BH mass is due to low amounts of accretion stemming from the effectiveness feedback has at heating the immediate environment around our sink particles, calling for an improvement of the feedback model at our resolution, one option being the inclusion of a fully coupled radiation hydrodynamical feedback (see \citet{Bieri+2017}).

Our results confirm that galactic encounters can trigger bursts of SF 
\citep[e.g.][]{BarnesHernquist1991,MihosHernquist1996,Springel+2005a,Gabor+2016}.
The first pericenter pass clearly increase the SF of both galaxies but
those increases are more evident beyond $\sim$ 500 pc from the galactic center,
when it reaches $\sim$ few $M_\odot/$yr. At these higher scales the SFR enhancement 
is due to the gas density increase triggered by the collision of the gaseous galactic 
spiral arms. Because the first pericenter pass has a nuclear separation of $\sim$ 2 kpc, 
most of the SF is localized at those distances from the center. In contrast, 
the second and third pericenter passes trigger bursts of SF at the inner hundred
of parsecs, again reaching $\sim$ few $M_\odot/$yr. At this stage the gas density has 
increased due to mass transport, resulting in a prominent nuclear SF burst.

Besides the SFR, the BHAR peaks also show correlations with pericenter encounters. 
Whereas one of the BHs has a growth rate correlated with its three pericenter 
passess the other one correlates better with its second and third pericenter passes. 
In both cases it is evident that the second and third pericenter passes
increase the BHAR, reaching values of $\sim 50-100\%$ and $\sim 25\%$ of the corresponding Eddington limits for the BHs (corresponding to a few $\sim\,10^{-2}\,M_\odot/$yr and a few $\sim\,10^{-3}\,M_\odot/$yr). Such 
high mass accretion rate onto the compact objects will trigger the AGN activity.

Both phenomena described above, i.e., star formation activity and BH accretion, are driven by the amount of gas available to form stars and to feed 
the BHs. Our simulation shows that pericenter passes correlate with peaks of gas mass
accretion rates driving gas mass density variations in the BH vicinity, i.e. inside 
its influence radius.
The first encounter produces a direct mass inflow of 
$\sim 3\,M_\odot/$yr outside of $\sim$ 500 pc, associated with the galaxy-galaxy crossing. 
This encounter triggers $\sim$ kpc scale SF in both galaxies. On the other hand, 
at smaller scales ($r\la$ 100 pc) the first pericenter produces a big increase in 
the mass accretion rate for one of the galaxies (reaching a short peak of $\sim 10\,M_\odot/$yr), and a smaller increase for the second one, but still enough to produce SF and to feed one of the 
BHs. The second and third pericenter passes produce a clear enhancement in mass accretion rate onto the nuclear galactic region. In fact, at the third closest passage the gas mass inflow at inner scales is simultaneously high for both systems and as such, the galactic 
gas entering the BH sphere of influence efficiently feeds the BHs and triggers nuclear 
SF bursts.
 
Neglecting magnetic fields and viscosity, any variation on the gas angular
momentum will be due to torques from both gravitational and pressure 
gradient forces (see appendix \ref{appD}). In other words, the merger 
triggers changes in the gas angular momentum due to variations in the gravitational
potential and gas pressure. The former are produced due to the dynamics of the merger 
which is characterized by strong gravitational interactions, and the latter is 
produced by gas layers with strong differences in density and/or temperature. 
Such conditions naturally arise when both galaxies cross each other and finally 
merge. We have shown that pericenter passes correlate with both gravitational and hydrodynamic torque peaks. 
In general, gravitational torques dominate over hydrodynamic torques but at inner scales
pressure gradient torques can reach values approaching that of the gravitational ones 
helping to radially transport gas in galactic disc. 
These torques redistribute angular momentum allowing inward mass transport
onto the galactic center. The high resolution of our simulation showed that such 
gas inflows can cross the BH influence radius producing peaks in the BHAR 
and triggering SF burst. 

\section*{Acknowledgements}
Powered@NLHPC: This research was partially supported by the supercomputing
infrastructure of the NLHPC (ECM-02). The Geryon cluster at the Centro de
AstroIngenieria UC was extensively used for the analysis calculations 
performed in this paper. JP is funded by ESO-Chile Comite Mixto grant ORP 
79/16. AE acknowledges partial support from the Center for Astrophysics 
and Associated Technologies CATA (PFB06) and Proyecto Regular Fondecyt 
(grant 1181663). G.C.P. acknowledges support from the University of Florida.

\begin{appendices}
\clearpage
\section{Appendix: Rotation center}
\label{appB}
The rotational center of a system composed by particles of mass $m_i$ at position 
$\vec{r}_i$ and acceleration $\vec{a}_i$, and where the amount of particles well-represent the phase space near such center, can be defined as the point 
$\vec{r}_{\rm rot}$ where the torque
\begin{equation}
\vec{\tau}_{\rm rot}=\sum_{i}\,m_i\,(\vec{r}_i-\vec{r}_{\rm rot})\times(\vec{a}_i-\vec{a}_{\rm rot})
\end{equation}
inside a given volume is null, with $\vec{a}_{\rm rot}$ the rotational center
acceleration. In a well approximated system which is supported by ideal rotation, all the accelerations 
will point to a common center, the rotational center, then the cross product 
position-acceleration will be null. In systems with a given degree of turbulence and strong noise in 
its acceleration field, such null point does not necessarily exist, here the task 
reduces to searching for minima in the torque field to define our rotational center, 
which necessarily introduces degeneracy in its estimation.

A kinematic approach to identify the rotational center of a system can be based in 
the previous dynamical definition. In this case, instead of focusing on the particle
accelerations it is useful to look at the particle velocities $\vec{v}_i$. Then the
rotational center will be the point where the angular momentum
\begin{equation}
\vec{L}_{\rm rot}=\sum_{i}\,m_i\,(\vec{r}_i-\vec{r}_{\rm rot})\times(\vec{v}_i-\vec{v}_{\rm rot})
\end{equation}
inside a given volume is maximized. Here $\vec{v}_{\rm rot}$ is the velocity of the
rotational center. Note that in this case the cross product position-velocity should 
be a maximum. As with the dynamical definition, if the system has a given degree of turbulence, it would be possible to find more than one center of rotation.
We note that if our context was understood as a generic dynamical system, our search criterion reduces to finding the best candidate fulfilling the characteristics of a non-stationary irrational vortex, where $\vec{L}_{\rm rot}$ is the local circulation field maxima.

The process of identifying a rotational center is computationally expensive as it requires
computing the angular momentum (or torque) inside a given volume for each point in 
the space. Given the 3D map for the modulus of the angular momentum it is necessary 
to look for peaks in the angular momentum distribution. In other words it is necessary 
to look for ``clumps'' of angular momentum. Given the ``clumps'' of angular momentum 
it is possible to compute the centroid of such objects to define rotational centers. 
Thus, identifying the stellar center of mass given an ansatz for the 
rotational center (the BH positions for instance) is faster, computationally.


\bigskip
\section{Appendix: Non-inertial frames}
\label{appC}
Inside an accelerating reference frame the Newtonian dynamical equations are modified.
In such moving frame an observer will describe the movement of any object as influenced 
by ``fictitious forces''. Quantitatively, from a moving system with position $\vec{R}$
with respect an inertial reference frame the force described by an observer at $\vec{R}$
acting on a particle at position $\vec{r}_i$ is
\begin{equation}
m_i\frac{d^2\vec{r}_i'}{d\,t^2}=\vec{F}_i-m_i\frac{d^2\vec{R}}{d\,t^2}-m_i\vec{\omega}\times(\vec{\omega}\times\vec{r}_i')-2m_i\vec{\omega}\times\vec{v}_i'-m_i\frac{d\,\vec{\omega}}{d\,t},
\end{equation}
where $m_i$ is the mass particle $\vec{r}_i'=\vec{r}_{i}-\vec{R}$ is the particle
position with respect the moving system position $\vec{R}$ and $\vec{r}_{i}$ the
particle position with respect an inertial reference frame. $\vec{F}_i$ is the net 
force acting on the particle $i$ (due to the magnetic, gravitational, viscous or
hydrodynamic contribution), $\vec{\omega}$ is the angular velocity of the 
moving system and $\vec{v}_i'=d\vec{r}_i'/dt$.

In the simple case when $\vec{\omega}=\vec{0}$, i.e. a moving reference frame without
rotation, with $\vec{a}_i'$ the particle acceleration with respect the moving system, 
$\vec{a}_{i}$ the particle acceleration with respect an inertial frame and $\vec{A}$ 
the moving system acceleration with respect the same inertial reference frame it is
possible to write 
\begin{equation}
\vec{a}_i'=\vec{a}_{i}-\vec{A},
\end{equation}

\section{Appendix: Torques-mass transport relation}
\label{appD}
The momentum conservation equation in its conservative form in 
Cartesian coordinates $x_i$ can be written as
\begin{equation}
\frac{\partial (\rho v_k)}{\partial t}+\frac{\partial}{\partial x_l}(R_{kl}+P_{kl}+B_{kl}-G_{kl}-S_{kl})=0,
\label{peq}
\end{equation}
where $\rho$ is the gas mass density and $v_i$ is the Cartesian component of 
the gas velocity. $R_{kl}$, $P_{kl}$, $B_{kl}$, $G_{kl}$ and $S_{kl}$ are 
the hydrodynamical stress, the pressure stress, the magnetic stress,
the gravitational stress and the viscous stress, respectively. The stresses 
are defined by: 
\begin{eqnarray}
R_{kl}&=&\rho v_k v_l,\\
P_{kl}&=&\delta_{kl}P,\\
B_{kl}&=&\frac{1}{4\pi}\left(B_k B_l-\frac{1}{2}B^2\delta_{kl}\right)\\
G_{kl}&=&\frac{1}{4\pi G}\left[\frac{\partial \phi}{\partial x_k}\frac{\partial \phi}{\partial x_l}-\frac{1}{2}(\nabla\phi)^2\delta_{kl}\right],\\
S_{kl}&=&\rho\nu\left(\frac{\partial v_k}{\partial x_l}+\frac{\partial v_l}{\partial x_k}-\frac{2}{3}\delta_{kl}\nabla\cdot \vec{v}\right),
\end{eqnarray}
where $P$ is the gas pressure, $B_k$ the cartesian component of the magnetic 
field, $B$ the modulus of the magnetic field, $\phi$ the gravitational 
potential, $\nu$ is the kinematic viscosity and $\delta_{kl}$ is the Kronecker 
delta symbol.

Neglecting the magnetic term and the dissipative-viscous term (Balbus 2003) the 
momentum conservation equation can be written as
\begin{equation}
\frac{\partial}{\partial t}(\rho v_k)+\frac{\partial}{\partial x_l}(\rho v_k v_l)+\frac{\partial P}{\partial x_k}-\rho\frac{\partial \phi}{\partial x_k}=0,
\label{momcon}
\end{equation}
and taking the cross product between the Cartesian position $\vec{x}$ and 
eq. \ref{momcon} (applying $\epsilon_{ijk}x_j$, with $\epsilon_{ijk}$ the 
Levi-Civita symbol) it is possible to derive the angular momentum 
conservation equation and after some algebra it is possible to write the 
$\hat{z}$ component of this equation as
\begin{equation}
\frac{\partial}{\partial t}\left(\rho\ell_z\right)=-\left[\ell_z\rho\,\nabla\cdot\vec{v}+\vec{v}\cdot\nabla(\rho\,\ell_z)+\tau_z^P+\tau_z^G\right],
\label{lz}
\end{equation} 
from where it is possible to get the gas mass density variation
\begin{equation}
\frac{\partial\rho}{\partial t}=-\frac{1}{\ell_z}\left[\rho\frac{\partial\ell_z}{\partial t}+\ell_z\rho\nabla\cdot\vec{v}+\vec{v}\cdot\nabla(\rho \ell_z)+\tau_z^P+\tau_z^G\right],
\label{mt}
\end{equation}
where $\ell_z=(\vec{x}\times\vec{v})\cdot\hat{z}$ is the $z$ component of 
the gas specific angular momentum, 
$\tau_z^G=\rho\,(\vec{x}\times\nabla\phi)\cdot\hat{z}$ is the $z$ component 
of the gravitational torque and 
$\tau_z^P=(\vec{x}\times \nabla P)\cdot\hat{z}$ is the $z$ component of 
the hydrodynamic torque.

Equation \ref{mt} relates the changes in gas density $\rho$ with torques 
$\tau_z^{P,G}$ acting on the gas. For a system starting from an 
axisymmetric stationary state with 
$\vec{v}=v(r)\,\hat{\theta}$ and $\rho=\rho(r)$ 
the azimuthal perturbations in both gas pressure and 
gravitational potential are the sources of changes in gas density, i.e. both 
hydrodynamic torques and gravitational torques are able to transport matter from 
a given radius to another radius of the system.

\end{appendices}

\end{document}